# Hemodynamic Bigger Hydrostatic Pressure Instead of Lower Shear Stress Aggravates Atherosclerosis

Brief title: Bernoulli's Fluid Equation in Atherosclerosis

Xinggang Wang [a, b], Junbo Ge [a, *]

[a] Department of Cardiology, Zhongshan Hospital, Fudan University. Shanghai Institute of Cardiovascular Diseases, Shanghai, China

[b] Institute of Biomedical Sciences, Fudan University, Shanghai, China

*Corresponding author

## ABSTRACT

The traditional view is that low shear stress of blood flow accelerates atherosclerosis. How does a low shear stress damage the vessel or aggravate atherosclerosis? It is a puzzling question. Furthermore, shear stress of veins is similar to that of Relatively Vulnerable Zones (RVZ) of arteries, while intimal thickening of veins is not obvious as the arteries. There are many places in arterioles with lower shear stress as the decrease of blood velocity, while the thickening of arterioles is not as obvious as that of the medium-sized arteries. These issues make it unconvincing for us to explain atherogenesis with lower shear stress. According to Bernoulli's fluid equation ($P+\frac{1}{2}\rho v^2+\rho gh=$constant), in a closed pipe full of flowing blood, at the same height, the hydrostatic pressure is inversely proportional to the second power of blood velocity. At any point of per unit mass of fluid micro cluster here, the reduction of $\frac{1}{2}\rho v^2$ would be converted into P. Therefore, hydrostatic pressure(P) is inversely proportional to $v^2$ in a very short distance or in the same transection of the artery. When blood micro cluster flows over a very short distance or the same transection of the artery, previous studies did not consider the conversion between $\frac{1}{2}\rho v^2$ and P. Therefore, low shear stress aggravates atherosclerosis is an appearance, and the essence is that these areas with smaller blood velocity have much bigger hydrostatic pressure, which aggravates atherosclerosis. This perfectly explains the predisposing sites of atherosclerosis.

**Main Text**

Atherosclerosis is prone to large and medium arteries which must bear much bigger mechanical forces, mainly hydrostatic pressure, shear stress and tensile stretch. In general, with gradual increase of branches and total sectional area, velocity and pressure of blood will gradually decrease from aorta to capillaries. However, local velocity and pressure of blood might also be different even in the same transection of artery for variations of vessel structure and location. Blood belongs to viscous fluid with certain viscosity in the body. In the large and medium arteries, blood velocity is so fast that viscoelasticity could be negligible. Therefore, the Bernoulli's equation could be applied to these arteries: $P+\frac{1}{2}\rho v^2+\rho gh$=constant or $P$=constant-$\frac{1}{2}\rho v^2$-$\rho gh$ (P: hydrostatic pressure, ρ: fluid density, v: blood velocity, g: gravitational acceleration, h: height). ρ and g are constants in an individual. The essence of Bernoulli's equation is energy conservation. At any point of per unit mass of fluid micro cluster, the sum of P, $\frac{1}{2}\rho v^2$ and ρgh is a constant. Even if the viscosity of blood is considered, the energy loss of blood flow should be very small over a very short distance (few centimeters, Fig.1). In addition, the energy loss of blood flow in the same transection is also very small due to the small diameter of blood vessel. At the same timepoint in a cardiac cycle, the constants of unit mass of fluid micro cluster are basically equal in a very short distance or in the same transection of artery, and Bernoulli's equation is still applicable here. At any point of per unit mass of fluid micro cluster here, the reduction of $\frac{1}{2}\rho v^2$ would be converted into P. Therefore, hydrostatic pressure(P) is inversely proportional to $v^2$ in a very short distance or in the same transection of the artery (Fig.1).

Different vessels have different susceptibility of atherosclerosis, and even in the same transection of the same artery, the susceptibility is different in different locations which is called eccentric lesions. Therefore, it is thought that this should be closely related to hemodynamics. We all know that the outer walls of bifurcations and the convex surface of curved arteries are usually the "Relatively Vulnerable Zones" (RVZ) where atherosclerosis is often more serious compared with their neighborhoods. For decades, many studies have confirmed that the areas of the outer walls of bifurcations have much smaller blood velocity than their neighborhoods(1). Therefore, "lower shear stress" is considered as a risk factor of atherosclerosis for decades. How does a lower shear stress damage the vessel or aggravate atherosclerosis? It is a puzzling question. Furthermore, shear stress of veins is similar to that of RVZ of arteries(1), while intimal thickening of veins is not obvious as the arteries. There are many places in arterioles with lower shear stress with the decrease of blood velocity, while the thickening of arterioles is not as obvious as that of the medium-sized arteries. These issues make it unconvincing for us to explain atherogenesis with "lower shear stress".

When blood micro cluster flows over a very short distance or the same transection of the artery, previous studies did not consider the conversion between $\frac{1}{2}\rho v^2$ and P. Of course, we did not mean to deny the previous studies, while these studies are very meaningful and made us to further consider its essence. RVZ are places where blood velocity is smaller than their neighborhoods. According to Bernoulli's equation (P+$\frac{1}{2}\rho v^2$+$\rho gh$=constant), RVZ have much bigger hydrostatic pressure than their neighborhoods. It is well known that hypertension with bigger hydrostatic pressure is the most common risk factor of atherosclerosis. This perfectly explains the predisposing sites of atherosclerosis, such as bifurcations and curved arteries (Fig.1). Therefore, "low shear stress" aggravates atherosclerosis is an appearance, and the essence is that these areas with smaller blood velocity have much bigger hydrostatic pressure, which aggravates atherosclerosis.

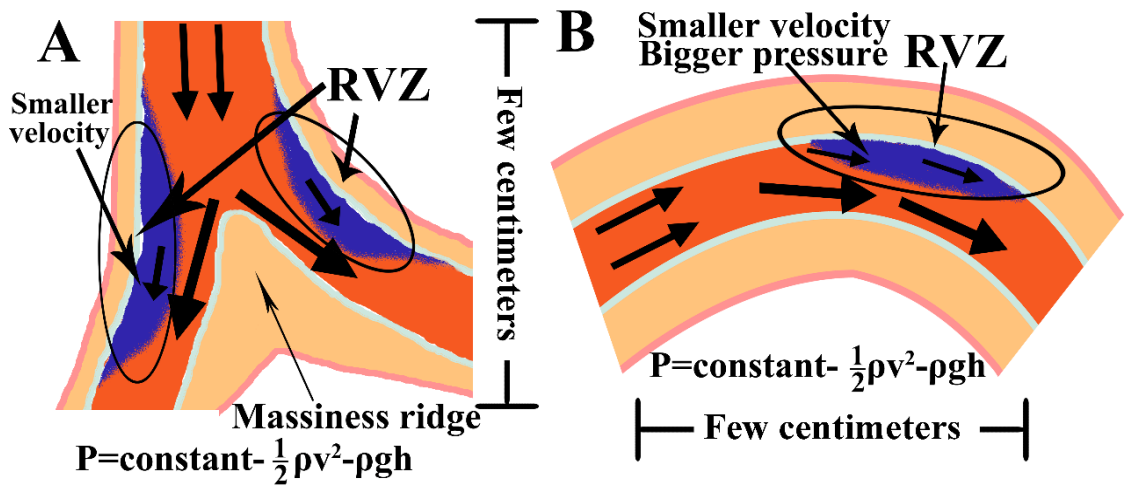

Fig.1 Hemodynamics in atherosclerosis

**(A)**：It is the illustration of bifurcations where and why atherosclerosis is prone to occur. Blood velocity near the outer walls of LAD/LCX bifurcations is much smaller than their counterparts in the same transection. With $P=constant-\frac{1}{2}\rho v^2-\rho gh$, these zones have much bigger hydrostatic pressure(P) near the outer walls of LAD/LCX bifurcations than their counterparts. This leads to more severe lesions in these "Relatively Vulnerable Zones" (RVZ).

**(B):** It is the illustration of curved artery where and why atherosclerosis is prone to occur. Blood velocity near the convex surface of curved arteries is smaller than that of others by knowledge of fluid physics of curved pipes. With $P=constant-\frac{1}{2}\rho v^2-\rho gh$, these zones have much bigger hydrostatic pressure(P) near the convex surface of curved arteries compared with other zones.

**Acknowledgements**

This work was supported by China Postdoctoral Science Foundation (no. 2018M641934) and a grant of Innovative Research Groups of the National Natural Science Foundation of China (81521001).

**Author contributions**

Junbo Ge supervised the study. Junbo Ge and Xinggang Wang designed the experiments. Xinggang Wang performed experiments, analyzed data, and wrote the manuscript. Junbo Ge made manuscript revisions.

**Competing interests**

There was no conflict of interest.